\documentclass[reprint,amsmath,amssymb,aps,superscriptaddress,longbibliography,preprintnumbers,prb]{revtex4-2}
\usepackage{graphicx}
\usepackage{dcolumn}
\usepackage{bm}
\usepackage[utf8]{inputenc}
\usepackage{amsfonts}
\usepackage{physics}
\usepackage{overpic}
\usepackage{color}
\usepackage[dvipsnames]{xcolor}
\usepackage[]{hyperref}
\usepackage{comment}
\usepackage{braket}
\usepackage{tikz}
\usetikzlibrary{tikzmark}
\usepackage[T1]{fontenc}
\usepackage{enumitem}
\usepackage{booktabs}
\usepackage{float}
\usepackage{dsfont}
\usepackage{blkarray, bigstrut}
\usetikzlibrary{tikzmark,calc,fit}
\usepackage{soul}
\usepackage[left]{lineno}

\usepackage{orcidlink}
\newcommand{\orcidalberto}{\orcidlink{0000-0002-6643-0738}}
\newcommand{\orcidnora}{\orcidlink{0000-0002-9490-2536}}
\newcommand{\orcidgianpaolo}{\orcidlink{0000-0002-4274-6463}}
\newcommand{\orcidsven}{\orcidlink{0009-0003-1603-0351}}
\newcommand{\orcidfabio}{\orcidlink{0000-0002-3429-8189}}
\newcommand{\orcidmarco}{\orcidlink{0000-0003-4647-8758}}
\newcommand{\orcidsimone}{\orcidlink{0000-0002-8882-2169}}
\newcommand{\orcidkarlo}{\orcidlink{0009-0008-0003-0986}}

\newcommand{\unipd}{Dipartimento di Fisica e Astronomia "G. Galilei" \& Padua Quantum Technologies Research Center, Universit{\`a} degli Studi di Padova, Italy I-35131, Padova, Italy}

\newcommand{\pdinfn}{INFN, Sezione di Padova, via Marzolo 8, I-35131, Padova, Italy}
\newcommand{\irb} {Institut Ruder Boškovi\'c, Bijeni\v{c}ka cesta 54, Zagreb 10000, Croatia}
\newcommand{\unizg} {University of Zagreb, Bijeni\v{c}ka cesta 30, Zagreb 10000, Croatia}

\begin{document}

\title{A new rung on the ladder: exploring topological frustration\\towards two dimensions}%

\author{Alberto Giuseppe Catalano\orcidalberto}
\affiliation{\unipd}
\affiliation{\pdinfn}
\affiliation{\irb}
\author{Nora Reinić\orcidnora}
\affiliation{\unipd}
\affiliation{\pdinfn}
\author{Gianpaolo Torre\orcidgianpaolo}
\affiliation{\irb}
\author{Sven Benjamin Kožić\orcidsven}
\affiliation{\irb}
\author{Karlo Delić\orcidkarlo}
\affiliation{\unizg}
\author{Simone Montangero\orcidsimone}
\affiliation{\unipd}
\affiliation{\pdinfn}
\author{Fabio Franchini\orcidfabio}
\affiliation{\irb}
\author{Salvatore Marco Giampaolo\orcidmarco}
\affiliation{\irb}

\date{\today}
             
\begin{abstract}


Topological frustration arises when boundary conditions impose geometric frustration in a quantum system, creating delocalized defects in the ground states and profoundly altering the low-energy properties. While previous studies have been concerned with one-dimensional systems, showing that the ground state structure can be described in terms of quasiparticle excitations, the two-dimensional setting remains unexplored. We address this gap by studying a three-legged antiferromagnetic quantum Ising ladder on a torus using tensor network methods, where topological frustration is induced by an odd number of spins along both spatial directions. Our results reveal the first instance in which topological frustration shifts the position of the quantum critical point. By studying the entanglement structure, we find that the ground state can be characterized as hosting three delocalized quasiparticles. This work builds the quasiparticle picture of topological frustration toward higher dimensions and more complex systems than those considered so far.



\end{abstract}

\preprint{RBI-ThPhys-2025-37}

\maketitle



\section{Introduction}
\label{sec:introduction}
In physics, we refer to the impossibility of simultaneously minimizing all local interactions in a system's Hamiltonian as \textit{frustration}. This often arises in magnetic systems, where the spins cannot be oriented in such a way that all pairwise interactions are satisfied at once. In these systems, the two main causes of frustration can be identified as the geometry or topology of the lattice, and the competition between different types of interactions (e.g., ferromagnetic vs. antiferromagnetic couplings).
Despite its negative connotation in everyday language, frustration has acquired a distinct significance in modern physics, where it is closely associated with rich and unconventional phenomena in classical and quantum many-body systems. Indeed, frustration plays a central role in the emergence of a variety of exotic quantum phases. One of the most intriguing is the quantum spin liquid phase~\cite{Savary2017}, a highly entangled magnetic state lacking long-range order even at zero temperature. Experimental realizations of such states have been reported~\cite{Balz2016, Gao2019, Knolle2019}, often involving materials with triangular~\cite{Wannier1950,Stephenson1970}, Kagome~\cite{Balents2010} or pyrochlore~\cite{Moessner1998_B,Moessner1998_L,Henley2005} lattice geometries that inherently promote magnetic frustration. 
Moreover, the recent development of quantum simulation platforms has enabled the direct emulation of frustrated spin chains and their dynamics, opening new avenues for experimentally probing non-equilibrium and quantum annealing processes in regimes otherwise inaccessible to classical computation \cite{King2021}.

\begin{figure*}[t]
    \centering\includegraphics[width=0.8\textwidth]{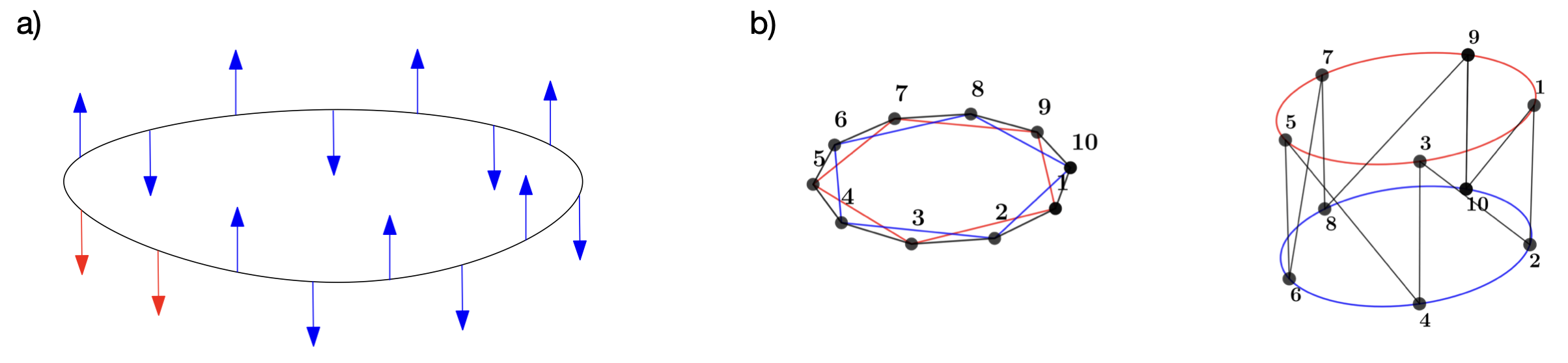}
    \caption{\textbf{Quantum spin models that exhibit topological frustration.} a) Nearest-neighbors AFM spin chain under FBCs. A system with odd number of sites can not avoid hosting at least one ferromagnetic defect (red pair of spins). The picture depicts one possible configuration. b) Interaction network of the ANNNI chain for $L=4n+2$ with $n\in\mathbb{N}$~\cite{Torre2025}.}
    \label{fig:Frustration1D}
\end{figure*}

In the past years, a form of weak frustration, dubbed topological frustration, has garnered a lot of interest in the context of one-dimensional spin-$1/2$ systems. Topological frustration (TF) was first introduced in the context of local antiferromagnetic (AFM) quantum spin chains by applying the so-called frustrated boundary conditions (FBCs)~\cite{Dong2016}, realized as a combination of periodic boundary conditions and odd number of spins. More generally, this kind of frustration is related to the presence of frustrated loops of extensive length, i.e., loops whose size diverges in the thermodynamic limit. In the classical limit, the presence of these frustrated loops produces a (super-)extensive ground state degeneracy. The addition of non-commuting terms in the Hamiltonian can lift these degeneneracies~\cite{Sen2008,Maric2022_fate,Catalano2022}, typically resulting in a gapless energy band. The emergence of such a band structure is therefore a pure consequence of TF, replacing the gapped phase which would otherwise exist in the same region of parameter space under different boundary conditions.
This produces several consequences in the low-energy physics of these systems, which are interesting both from a theoretical and an application point of view. Indeed, from one side, it has been shown that the presence of TF can destroy local order parameters~\cite{Maric2020_destroy}, induce novel phase transitions~\cite{Maric2020_induced} or change the nature of already existing ones~\cite{Maric2022_nature,Sacco2023}. This enrolls TF systems to the list of those escaping the standard Landau paradigm of phase transitions~\cite{Landau1965}. On the other hand, TF also produces an enhancement of the quantum resources of the ground states of 1D spin chains, measured, for example, by the entanglement entropy or the non-stabilizerness~\cite{Giampaolo2019,Odavic2023}. Such features can be exploited for quantum technological applications~\cite{Catalano2024_battery}. 
Interestingly, most of the effects of TF can be efficiently described within a quasiparticle picture, corresponding to the presence of delocalized excitations induced by TF in the ground state of 1D spin chains~\cite{Giampaolo2019, Torre2025}.

While the physics of TF has been studied on one-dimensional examples up to now, the goal of this work is to extend the investigation towards two-dimensional systems. The challenge turns out to be a very complicated one not only from the analytical, but also from the numerical point of view. While in 1D systems the degeneracy of the classical ground state manifolds typically scales polynomially with the system's size, the number of degenerate classical ground state configurations can become exponential in 2D TF systems. This results in a very high density of states in the low-energy bands, which increases the complexity of simulation with most of the state-of-the-art numerical algorithms. Accordingly, in this manuscript we take a first step towards the study of full two-dimensional TF systems, and consider a quantum Ising model on a three-legged ladder with the topology of a torus. Interestingly, such geometry induces an extensive amount of frustration with respect to the total system size, which is not the case for a chain or a fully 2D system. Using tensor network numerical techniques, we show that the presence of TF modifies the phase diagram of the model by shifting the critical point position. Moreover, we show that the bipartite entanglement entropy of this system is compatible with the presence of three bosonic excitations in the ground state of the system.

The manuscript is organized as follows. In Section~\ref{sec:recap} we review the existing results on quasiparticle picture emerging in one-dimensional systems and explain the nature of the excitations introduced by TF. Then, in Section~\ref{sec:theory} we define the model under investigation and analyze its phase diagram by studying the behaviour of the energy gap and the bipartite entanglement entropy. Focusing on the latter quantity, in Section~\ref{sec:quasiparticles} we extend the quasiparticle picture introduced in 1D systems to our model, and finally draw our conclusions in Section~\ref{sec:conclusions}.

\section{The quasiparticle picture in 1D}
\label{sec:recap}

The simplest systems that allow us to understand why TF induces quasiparticle excitations in spin systems are one-dimensional quantum spin chains with nearest neighbors interactions. The prototypical model within this family is the fully anisotropic Heisenberg chain, a.k.a. the XYZ chain, whose Hamiltonian reads
\begin{equation}
    \label{eq:XYZ}
H_{\mathrm{XYZ}}=\sum_{\alpha=x,y,z}J_\alpha\sum_{\ell=1}^L\sigma^\alpha_\ell\sigma^\alpha_{\ell+1}+h\sum_{\ell=1}^L\sigma^z_\ell,
\end{equation}
where $h$ is an external magnetic field along the $\vec{z}$ axis, while $\vec{J}=(J_x,J_y,J_z)$ encodes the spin-spin couplings along each spatial direction. When the dominant interaction in Eq.~\eqref{eq:XYZ} is AFM, the application of FBCs induces TF in the system. In this case, the frustrated loop coincides with the whole chain (see Fig.~\ref{fig:Frustration1D}a), and thus has extensive length. To understand the emergence of quasiparticle excitations, one could follow a perturbative approach. Indeed, at the classical point $h=J_y=J_z=0$ the ground state manifold of the system contains $2L$ degenerate classical configurations, corresponding to the $L$ possible locations of the ferromagnetic defect, with a factor $2$ coming from the $\mathbb{Z}_2$ symmetry of the model. Adding a small perturbation, e.g., $h\neq0$, lifts the degeneracy and, when the system is translationally invariant, the ground state is given by a state in which the ferromagnetic defect is completely delocalized along the chain. This corresponds to the injection of a quasiparticle excitation in the ground state of the system, which is reflected by the peculiar structure of the bipartite entanglement entropy. In the thermodynamic limit, it was indeed shown that the entanglement entropy $S$ can be expressed as~\cite{Giampaolo2019}

\begin{equation}
    S(x)= S^{QP}(x)+S_{loc}(h,\vec{J})+\log_2 2.
\label{eq:ee-decomposition1}
\end{equation}
where $x=l/L$ is the ratio between the length $l$ of the chosen subpartition and the total system size. The contribution $S_{loc}(h, \vec{J}) + \log_2 2$, where $S_{loc}$ encodes the contribution of local quantum correlations to the entanglement entropy, turns out to correspond to entanglement entropy of the corresponding unfrustrated model (i.e., the one obtained switching to ferromagnetic interactions), and
\begin{equation}
    S^{QP}(x)=-x\log_2 x -(1-x)\log_2(1-x) \equiv q(x)
\label{eq:one-excitation}
\end{equation}

\noindent corresponds to the entanglement contribution of the excitation induced by TF. Here, $q(x)$ can be interpreted as the entropy of the classical probability of finding the particle inside the bipartition. Because of this extra contribution, $S^{QP}(x)$, to the entanglement entropy, the TF ground states violate a strict area law, developing a finite (non-divergent) dependence on the size of the bipartition~\cite{Giampaolo2019}.

Similar results have also been obtained in systems with more complex interactions beyond nearest neighbors ones. This is the case, for example, of the axial next-to-nearest-neighbors Ising (ANNNI) chain, described by the Hamiltonian 
\begin{equation}
    \label{eq:ANNNI}
H_{\mathrm{ANNNI}}=J_1\sum_{\ell=1}^L\sigma^x_\ell\sigma^x_{\ell+1}+J_2\sum_{\ell=1}^L\sigma^x_\ell\sigma^x_{\ell+2}+h\sum_{\ell=1}^L\sigma^z_\ell.
\end{equation}
Here, $\vec{J} = (J_1, J_2)$ are the nearest- and next-to-nearest-neighbour interactions along the $x$ direction, and $h$ is again the external field along $z$. For $J_2>0$ this model has an extensive amount of frustration, since the nearest and next-to-nearest preferred alignments cannot be simultaneously fulfilled. Moreover, in the so-called antiphase of the model, i.e., for $J_2>J_1/2$, for $L=4m+2\wedge m \in \mathbb{N}$ the system possesses two interacting TF loops, each one of length equal to $L/2$ (see Fig.~\ref{fig:Frustration1D}b). The entanglement entropy can be decomposed in a similar way as Eq.~\eqref{eq:ee-decomposition1} also in this case. In the thermodynamic limit it reads~\cite{Torre2025}:

\begin{equation}
    S(x)= S^{QP}(x)+S_{loc}(h,\vec{J})+2\log_2 2,
\end{equation}
where again the term $S_{loc} (h, \vec{J}) + 2\log_2 2$ is equivalent to the entanglement entropy of the corresponding unfrustrated model, and $S^{QP}(x)$ is the contribution due to the topological excitations 
\begin{equation}
    S^{QP}(x)= q(y^+(x))+q(y^-(x)),
\end{equation}
with $y^{\pm}(x)=x\pm\sin(\pi x)/\pi$. Interestingly, in this case $S^{QP}(x)$ corresponds to the entanglement entropy of two free fermions on a ring~\cite{Torre2025, Berkovits2013}, each having probability $y^{\pm}(x)$ of being in the chosen subpartition. The asymmetry in these two probabilities, respectively enhanced or depleted by a factor $\sin(\pi x)/\pi$, is a direct consequence of the Pauli exclusion principle. 

Therefore, to sum up, we would like to highlight that a clear structure emerges in the analytical expression of the entanglement entropy of TF systems, which can be expressed through contributions of three terms:
\begin{enumerate}
  \item a term $S_{loc}(h,\vec{J})$ related to local correlations in the systems and, in the thermodynamic limit, this is the same as that observed in unfrustrated systems~\cite{Giampaolo2019}, modulo the constant term discussed next;

  \item a constant term related to the symmetries of the model (e.g., $\mathbb{Z}_2$ or $\mathbb{Z}_2\times\mathbb{Z}_2$ for the Ising and ANNNI chain respectively), which also coincides with the logarithm of the number of eigenvalues of the reduced density matrix of the quasiparticle model associated to $S^{QP}(x)$ in the thermodynamic limit. Such term is essentially the topological entanglement entropy term introduced in~\cite{Kitaev2006}. In previous studies of the entanglement entropy in TF systems, this constant term coincided between the frustrated and the non-frustrated counterpart. In this work, instead, we will show that the two cases are different;
  
  \item 
  what is left is a term $S^{QP}(x)$ dependent on the ration between the length of the chosen subpartition and the total length of the chain, which typically highlights the presence and the nature of quasiparticle excitations in the ground state.

\end{enumerate}


\section{Topologically frustrated Ising model on a ladder}
\label{sec:theory}

\begin{figure*}[t]
    \includegraphics[width=0.9\textwidth]{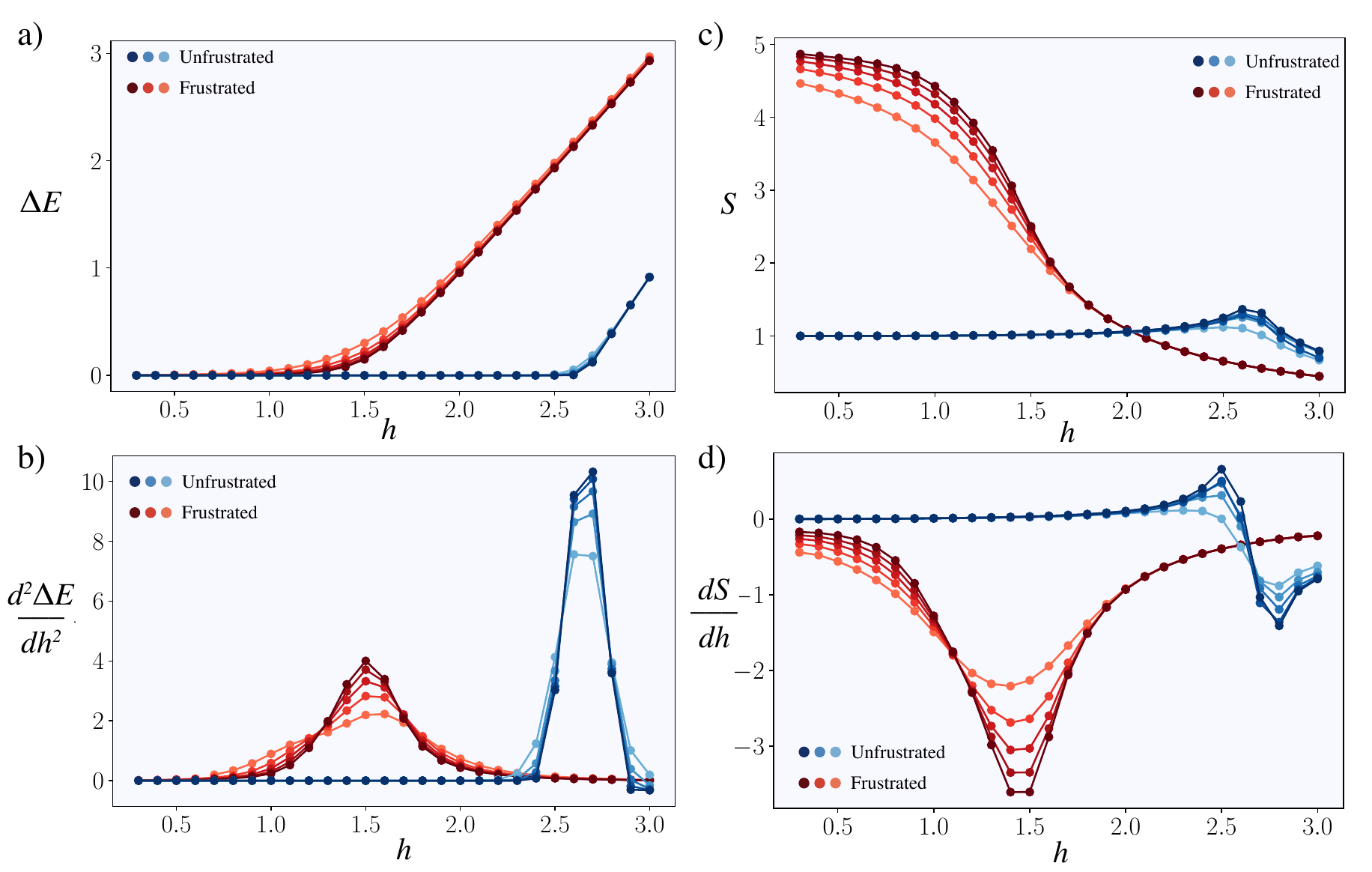} 
  \caption{\textbf{Phase diagram of the quantum Ising model on an L$\times$3 lattice}. a) The energy gap $\Delta$ between the ground and the first excited state and b) its second derivative as a function of external field $h$. c) The entanglement entropy $S$ between the system's halves and d) its first derivative as a function of external field $h$. In all panels, the blue curves correspond to the unfrustrated model (FM with PBC), and the red curves correspond to the frustrated model (AFM with PBC). For both unfrustrated and frustrated case, the curves correspond to $L=21,31,41,51,61$ in $L\times 3$ lattice, from the lightest to the darkest color respectively.}
  \label{fig:ph-diag}
\end{figure*}

Here, we consider a quantum Ising model in a transverse field on a three-leg ladder $L \times 3$ with periodic boundary conditions along both directions, i.e., a thin torus:

\begin{equation}
\begin{split}
   \hat{H} &=  \sum_{i,j=1}^{L_x, 3} (J_x\sigma^x_{i,j} \sigma^x_{i+1,j}+J_y\sigma^x_{i,j} \sigma^x_{i,j+1}
            - h\sigma^z_{i,j}),
\end{split}
\label{eq:ising-ham}
\end{equation}

\noindent where $\vec{J} = (J_x, J_y)$ parametrize the nature and strength of the interaction along the two spatial directions, and $h>0$ is the external field strength. Although a complete analytical description of the model's phase diagram is not available, numerical studies have shown that, in the purely 2D ($L \times L$) isotropic ($J_x=J_y$) case a quantum phase transition between an ordered phase (in which the specific order is determined by the sign of $J$) and a disordered phase (with the spins aligned along the direction of the external field) is expected when $h/J\approx3.04438$~\cite{Blote2002, Schmitt2022}. 

TF in this model is induced as a combination of odd lattice sites number ($L= 2m+1 \wedge m \in \mathbb{N}$), periodic boundary conditions (PBC), and antiferromagnetic (AFM) interactions, i.e., $J_x, J_y > 0$. In the analyses that follow, we alternate between the frustrated and unfrustrated cases by switching between AFM and ferromagnetic (FM) interactions. We stress that, by changing the sign of the interactions or by closing the boundary conditions, TF induces an extensive amount of frustration in the system. This is a peculiar feature of ladder systems and sets such models apart from the ANNNI model discussed in the previous section, where boundary conditions controlled only an intensive number of frustrated loops. In the rest of this manuscript we fix $|J_x|=1$ and consider $J_x=J_y$, unless explicitly stated otherwise. We carry out a numerical analysis of the frustrated and unfrustrated quantum Ising model on $L \times 3$ lattices using the density matrix renormalization group (DMRG) algorithm for matrix product states (MPS). See Appendix \ref{sec:numerical-method} for details of the numerical method and convergence parameters.


\subsection{Quantum phase diagram}\label{sec:qpd}

Since this is an Ising model, it is characterized by a $\mathbb{Z}_2$ symmetry, represented by the global magnetization parity $\prod _{i_x, i_y} \sigma^z_{i_x,i_y}$. We compute the lowest-energy state in both even and odd symmetry sector, giving us access to the ground and the first excited state energy, i.e., the energy gap. Moreover, we compute and analyze the entanglement entropy along different bipartitions.  We characterize the phase diagram by analysing the behaviour of the energy gap and the entanglement entropy for a transversal cut separating two halves of the torus for different external field values. 

The energy gap and its second derivative as a function of the external field are shown in Fig.~\ref{fig:ph-diag}a-b. The gap opening, i.e., the jump in its second derivative signals the presence of the second-order quantum phase transition. Comparing the unfrustrated and frustrated cases, we observe a clear shift of the quantum critical point position $h_c$ in the presence of TF. Although it is well known that TF can alter the nature of quantum phase transitions or induce novel ones, this represents the first instance of a critical point shift in this class of systems.

\begin{figure}[t!]
  \centering
    \includegraphics[width=0.9\columnwidth]{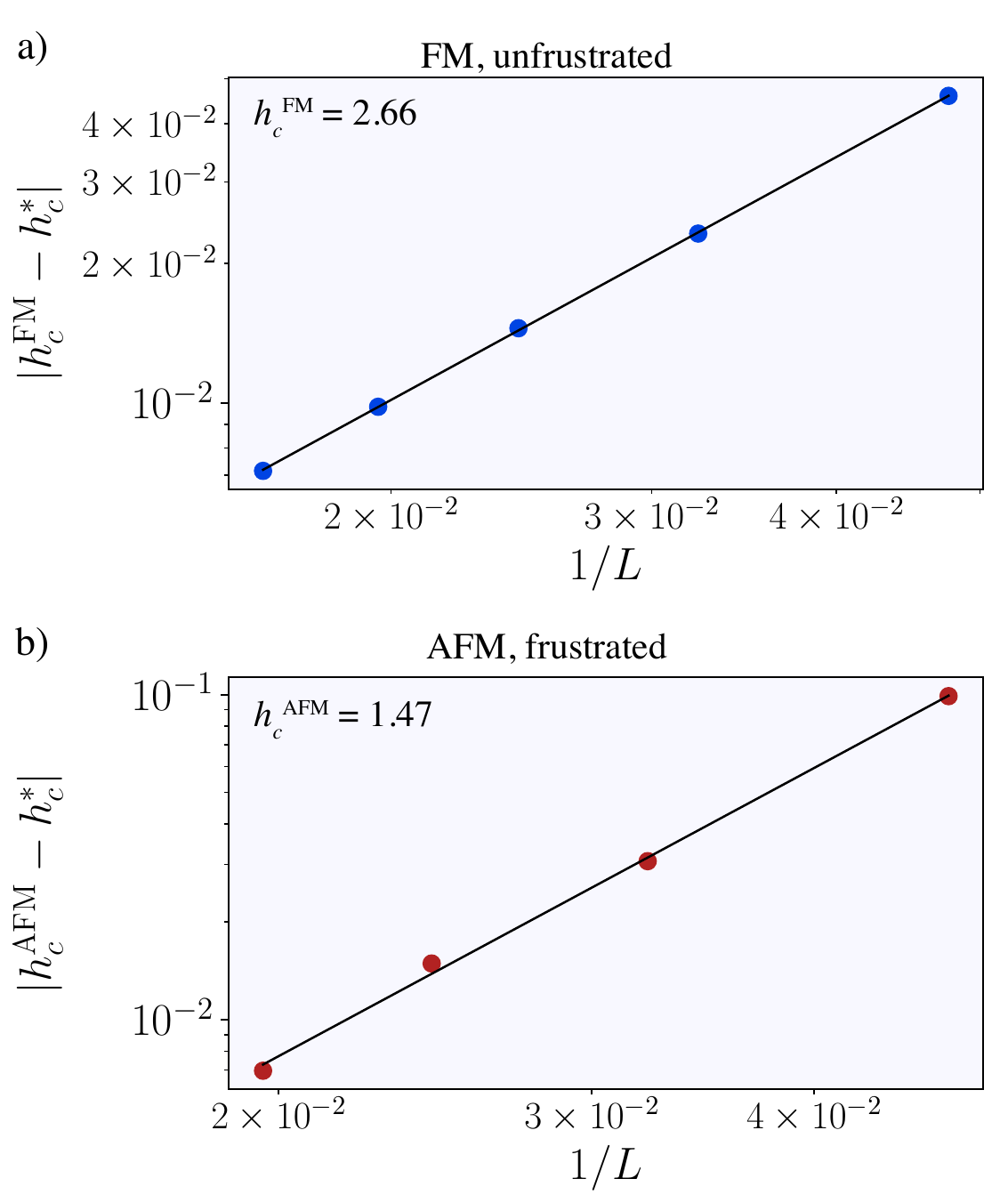} 
  \caption{\textbf{Extrapolation of quantum critical point values in quantum Ising model on an L$\times$3 lattice}. a) Difference between the finite-size critical points $h_c^*$ and the extrapolated thermodynamic value $h_c^{FM}$ as a function of inverse size $1/L$ for the unfrustrated case (FM with PBC). The finite-size critical points $h_c^*$ are extracted as the maxima of entanglement entropy. b) Difference between the finite-size critical points $h_c^*$ and the extrapolated thermodynamic value $h_c^{AFM}$ as a function of inverse size $1/L$ for the frustrated case (AFM with PBC). The finite-size critical points $h_c^*$ are extracted as the minima of the first derivative of entanglement entropy. In both panels, the black lines mark a polynomial fit, showing the convergence to a corresponding critical point in the thermodynamic limit. The obtained fitting coefficients are reported in Appendix~\ref{app:fitting}.}
  \label{fig:extrapolation}
\end{figure}

Fig.~\ref{fig:ph-diag}c shows entanglement entropy and Fig.~\ref{fig:ph-diag}d its first derivative as a function of external field. As expected from the results previously obtained in 1D systems, the behaviour of the entanglement entropy in the frustrated case differs with respect to the non-frustrated case. While the unfrustrated system obeys an area law and entanglement entropy quickly saturates for moderate system sizes for $h\ll h_c^{AFM}$, for the same external fields the states in the frustrated case are highly entangled. As discussed in Sec.~\ref{sec:recap}, this behaviour is typically explained in terms of the additional contributions coming from the delocalized excitations induced by TF. Moreover, we also observe that while in the non-frustrated case the quantum phase transition is directly signaled by the peak in entanglement entropy, this is not the case for the frustrated case in Fig.~\ref{fig:ph-diag}c, for which the entanglement entropy is showing smooth behaviour. However, we observe that the quantum phase transition in frustrated system manifests as a peak in the first derivative of the entanglement entropy (Fig.~\ref{fig:ph-diag}d). The positions of quantum critical points detected by entanglement entropy are in agreement with those detected by the energy gap. We estimate the position of critical points in the thermodynamic limit to be $h_c^{FM} = 2.660 \pm 0.001$ for the unfrustrated system (Fig.~\ref{fig:extrapolation}a) and $h_c^{AFM} = 1.47 \pm 0.02$ for the frustrated system (Fig.~\ref{fig:extrapolation}b), extrapolated from the entanglement entropy and its first derivative.


Both critical points are different from the one observed in a fully 2D system due to the small finite size in the vertical direction of the ladder, i.e., $L_y=3$. This phenomenon was also observed in other 2D geometries in which the thermodynamic limit is taken only along one of the two directions, such as infinite cylinders with a finite base ring size~\cite{Hashizume2022}. However, to the best of our knowledge, this is the first time that a shift in the critical point is observed switching from AFM to FM interactions or, equivalently, switching from standard periodic to frustrated boundary conditions.
This shift in the phase transition point can be explained heuristically in the following way. In an $L_x\times L_y$ system, the presence of TF introduces $L_x+L_y$ ferromagnetic defects which increase the energy contribution from the nearest-neighbors term with respect to the non-frustrated case. For an $L \times L$ system, the number of defects is thus $2L$, which is subleading to the total number of spins $L^2$ and therefore the effect is subdominant in the thermodynamic limit. However, for an $L \times 3$ system, the number of ferromagnetic defects, $L + 3$, is comparable to the total number of spins $3L$. Therefore, the quasi-staggered order with ferromagnetic defects remains less stable also in the thermodynamic limit, and hence a smaller magnetic field is sufficient to induce a transition to a disordered phase compared to the non-frustrated case. More in general, one would thus expect a similar behavior whenever changing the boundary conditions affects, in the thermodynamic limit, the ration between the total system size and the number of bonds in the system. Let us remark once more that, although not surprising, this is the first reported instance in which TF, that is frustration due to boundary conditions, is extensive and able to move the phase transition point. 

\subsection{Scaling of energy gaps}\label{sec:qpd_gap}
\begin{figure}[t!]
  \centering
    \includegraphics[width=\columnwidth]{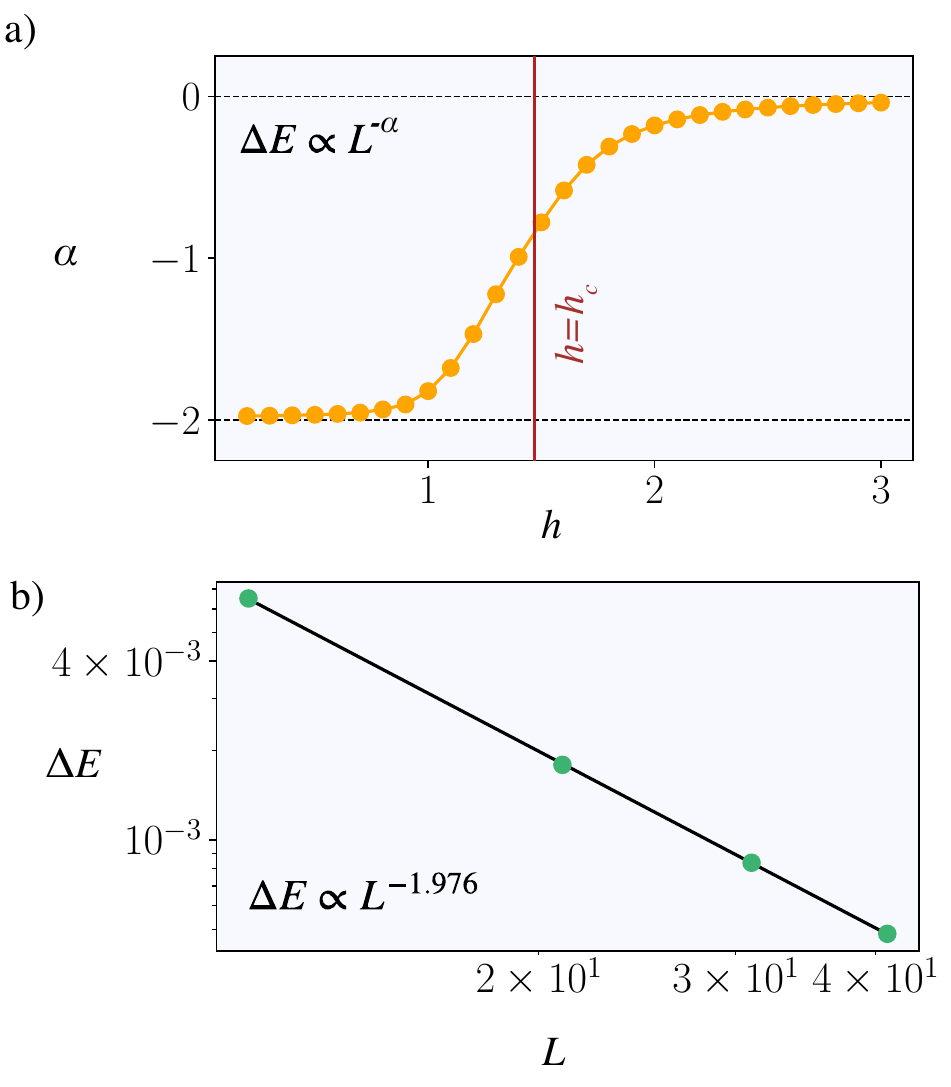} 
  \caption{\textbf{Scaling of energy gap with system size for frustrated quantum Ising model for L$\times$3 lattice}. a) The plot shows the scaling coefficient $\alpha$ fitted on a polynomial function $\Delta E \propto L^{-\alpha}$, as a function of external field $h$. Error bars are plotted, but not visible because they are several orders of magnitude smaller than the scaling exponent values. b) The energy gap $\Delta E$ at $h=0.2$ as a function of lattice side size $L$ plotted together with the fitted line with the scaling exponent $\alpha= 1.976$.}
  \label{fig:scalings}
\end{figure}
To further characterize the topologically frustrated system, we study the scaling of the energy gap with the system size at different points of the phase diagram. We fit the gap values for different lattice sizes to a polynomial function $\Delta E \propto L^{-\alpha}$, $L$ being the length of one side of the $L\times$3 lattice and $\alpha$ being the fitted scaling exponent. The resulting scaling exponents and the corresponding fit errors are shown in Fig.~\ref{fig:scalings}a). Errors bars are not visible because they are several orders of magnitude smaller than the scaling exponent. We can clearly distinguish between three different regions corresponding to the two phases and the critical region around quantum critical point.

In the $h\gg h_c$ region, the leading contribution to the Hamiltonian comes from the external field and thus this phase is not influenced by TF. Therefore, the energy gap approaches $\Delta E \propto const.$ in the thermodynamic limit which corresponds to the energy penalty of a one spin flip. Accordingly, the resulting scaling coefficient in Fig.~\ref{fig:scalings}(a) approaches $\alpha=0$ in the limit of strong external field. In the topologically frustrated phase, i.e., $h\ll h_c$ region, the scaling exponent approaches $\alpha= 2$, i.e., $\Delta E \propto L^{-2}$ in the limit of small $h$. The inverse quadratic scaling of energy gap $\Delta E$ as a function of system size $L$ is shown in Fig.~\ref{fig:scalings}(b) for $h=0.2$. Although there is no analytical solution available for our model, it is interesting to point out that this scaling of the gap is compatible with the 1D topologically frustrated Ising chain. In that case, the model is integrable, and the scaling of the gap is a direct consequence of the single-particle dispersion relation and the quantization of lattice momenta. The region between the two phases shows a transition in the value of the scaling exponent. We observe $\alpha \approx 0.82$ in the vicinity of the quantum critical point, compatible with the expected $\alpha=1$ conformal invariance scaling.

\section{Quasiparticle picture}\label{sec:quasiparticles}

In analogy with the work carried out for one-dimensional spin chains, here we provide a quasiparticle interpretation for the physics of the topologically frustrated system, in particular for the behaviour of the entanglement entropy. 

\begin{figure}[H]
  \centering
    \includegraphics[width=0.4\textwidth]{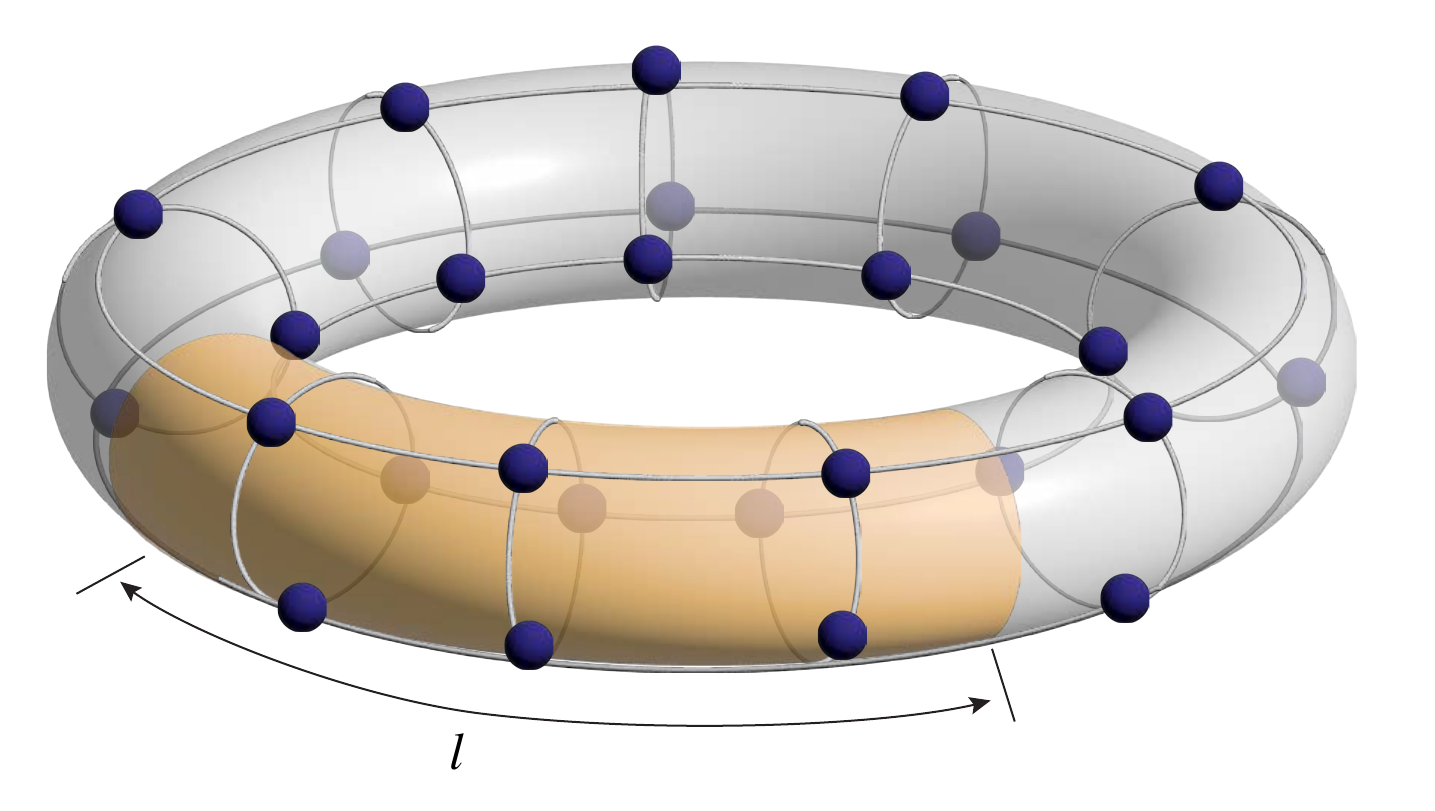} 
  \caption{\textbf{Choice of the subpartitions of entanglement entropy.} The topology of an $L\times3$ lattice with periodic boundary conditions corresponds to a thin torus. For the computation of entanglement entropy, we choose the bipartitions corresponding to the tubes of length $l$. The torus is depicted for $L=9$ and the orange area highlights an example of the tube of length $l=3$.}
  \label{fig:torus}
\end{figure}

\begin{figure*}[t]
    \includegraphics[width=1\textwidth]{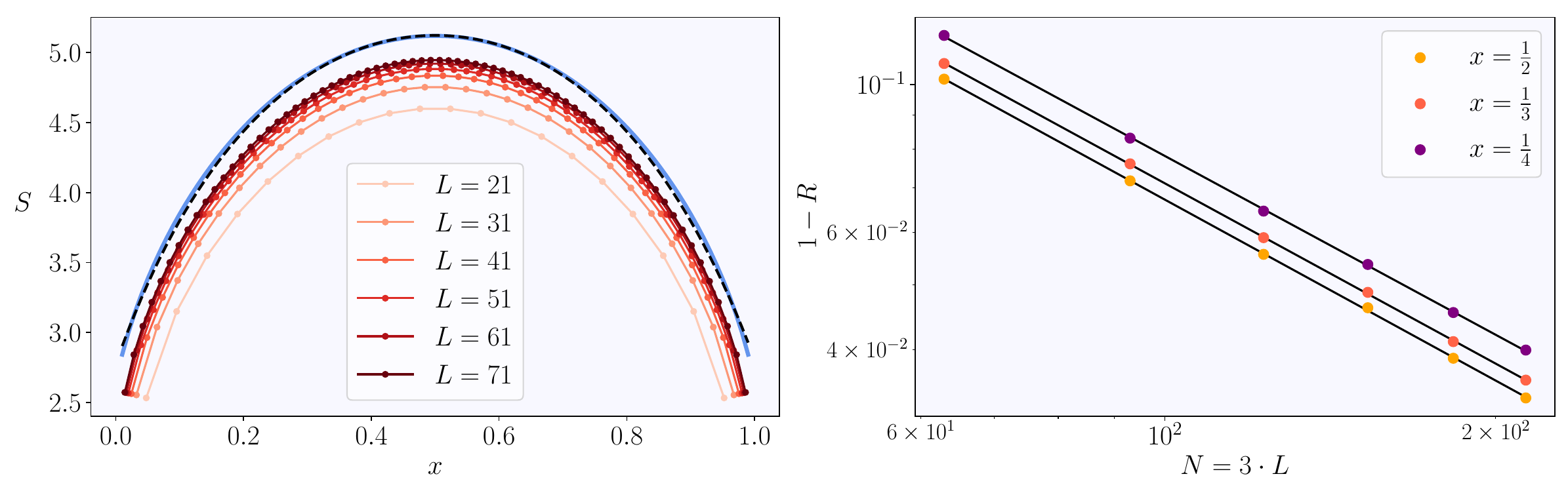} 
  \caption{{\textbf{The quasiparticle picture of entanglement entropy in topologically frustrated quantum Ising model on an $L \times 3$ lattice}. a) Bipartite entanglement entropy $S(x)$ as a function of the bipartition size $x$. Red points represent numerical data for the specified sizes $L$. The black dashed line represents the extrapolated values in the thermodynamic limit and the solid blue line represents the quasiparticle ansatz for a system of one independent particle and two bosons in the thermodynamic limit. b) The convergence of our data towards the quasiparticle ansatz, $1-R$, plotted as a function of the total system size $N=3 \cdot L$ for different bipartition sizes $x$. Black lines show the polynomial fit, with the obtained fitting coefficients reported in Appendix~\ref{app:fitting}.}}
  \label{fig:QP_picture}
\end{figure*}

Accordingly, we numerically study the bipartite entanglement entropy along different cuts of the system. The topology of an $L\times 3$ lattice with periodic boundary conditions corresponds to $3$ rings of $L$ sites connected to form a torus as in Fig.~\ref{fig:torus}. We choose the bipartition cuts such that each subsystem consists of a tube made by $l$ adjacent small rings of $3$ sites each, highlighted in orange in Fig.~\ref{fig:torus} in the example of $l=3$. Within this setting, we study the dependence of the entanglement entropy on the bipartition ratio $x=l/L$. The results of this analysis are shown in Fig.~\ref{fig:QP_picture}.

In particular, we find that the entanglement entropy obtained numerically is very well reproduced by 

\begin{equation}
\begin{aligned}
S(x,h,\vec{J}) &= S^{QP}(x) + S_{\text{loc}}(h,\vec{J}) + c, \\[0.5em]
S^{QP}(x) &= q(x) + b(x), \\
q(x) &= -x\log_2 x - (1-x)\log_2(1-x), \\
b(x) &= 2q(x) - 2x(1-x),
\end{aligned}
\label{eq:QP_ansatz}
\end{equation}

\noindent with $c$ as a constant. Here, $q(x)$ is the entanglement entropy of a single particle state as in Eq.~\eqref{eq:one-excitation} and $b(x)$ is the entanglement entropy of the ground state of a system of two free bosons hopping on a ring of $L$ sites. In analogy to Ref.~\cite{Castro-Alvaredo2018}, the expression of $S^{QP}(x)$ in Eq.~\eqref{eq:QP_ansatz} corresponds to the entanglement entropy of a state with three bosons, two of which occupy the same quantum state and the third occupying a different one. 

The term $b(x)$ differs from the entanglement entropy of two fully independent quasiparticles, i.e., $2q(x)$, because of the quantity $-2x(1-x)$. This extra contribution stems from the fact that two bosons, in their ground state, are indistinguishable, as reflected by the system's reduced density matrix which possesses only three non-zero eigenvalues, i.e.,
\begin{equation}
    \begin{cases}
        \lambda_0= (1-x)^2\\
        \lambda_1=2x(1-x)\\
        \lambda_2=x^2
    \end{cases}.
\end{equation}
For distinguishable particles, e.g., two bosons with different momenta, the spectrum of the reduced density matrix would be degenerate, with
\begin{equation}
    \begin{cases}
        \lambda_0= (1-x)^2\\
        \lambda_1=x(1-x)\\
        \lambda_2=x(1-x)\\
        \lambda_3=x^2
    \end{cases},
\end{equation}
yielding a total entanglement entropy of $2q(x)$.

We tested the compatibility between this ansatz and the numerical results obtained using DMRG by computing the ratio
\begin{equation}
    R=\frac{S^{\mathrm{DMRG}}(h,\vec{J},x)-S_{loc}(h,\vec{J})-c}{S^{QP}(x)},
\end{equation}
for different values of the bipartition size,
and the results are shown in Fig.~\ref{fig:QP_picture}b. As we stated earlier in Sec.~\ref{sec:recap}, in the thermodynamic limit the contribution to the entanglement entropy coming from local correlations far from critical points is the same in the frustrated and non-frustrated case. Hence, to isolate and extract the value of $S_{loc}(h,\vec{J})$ we computed the entanglement entropy of the non-frustrated (ferromagnetic) model and substracted the constant $\log_2 2$ term associated to its $\mathbb{Z}_2$ symmetry. As it turns out, anyways, for small values of the external field $h$, this quantity is very small, of the order of $10^{-10}$. The numerics support a power law convergence of $R$ to $1$ in the limit $L\to\infty$, and hence the validity of the quasiparticle picture proposed in Eq.~\eqref{eq:QP_ansatz}. Extrapolating the results to $L\to\infty$ we find a value of $c=2.62$. The value obtained for $c$ is very close to $\log_2 6$, i.e., to the logarithm of the number of non-zero eigenvalues of the reduced density matrix of the system of three bosons depicted in our ansatz. 
Notice that the constant $c$ turns out to be different between the frustrated and non-frustrated case, due to a different degeneracy of the entanglement spectrum.

To elucidate the physical content of our results, we could start by considering the limit case of $J_y=0$, in which the system is reduced to $3$ independent rings each of length $L$. In this case, in the thermodynamic limit, we know exactly that the bipartite entanglement entropy of the system will be that of three independent particles (we simply multiply the results obtained for a single Ising chain times the number of independent rings, i.e., $S^{QP}_{ind}=3q(x)$). 

Now, if we instead turn on the interaction $J_y>0$ at the classical point ($h=0$), following the derivation of the classical ground state manifold given in Appendix~\ref{ap:classical_gs_manifold} we observe that, because of the interaction constraints imposed by $J_y$, one of the three rings is still able to choose its configuration `freely', while the two remaining rings will be constrained by an effective interaction between their kinks. Moreover, for the system sizes considered here, at most two rings can possess a kink in the same position. These observations can then be translated into the picture of three bosons, two of which occupying the same state, described in Eq.~\eqref{eq:QP_ansatz}. 

Finally, it is worth mentioning that this picture is also reminiscent of some known results for AFM spin ladders in absence of frustration. In the case of an odd number of legs, it is known that the ground state of such systems can be described in terms of a `free' spin-1/2 ring and a singlet state realized by the remaining legs~\cite{Giamarchi2004}. 

Other quasiparticle interpretations of the entanglement results, such as three free fermions, three free bosons in the same state or three independent rings fail to describe the numerical results, as reported in Appendix~\ref{ap:ansatze}.


\section{Conclusions} \label{sec:conclusions}
In this work, we extended the study of topological frustration to a quasi-2D setting by analyzing a three-legged quantum Ising ladder with a toroidal topololgy. In this way, we provide a concrete starting point for future studies on topological frustration in fully two-dimensional models, where exponentially growing ground state degeneracy and low-energy density of states pose substantial analytical and computational challenges, but also opportunities for a rich phenomenology.

Our numerical results demonstrate that topological frustration has a quantitative effect on the phase diagram, shifting the critical point relative to its unfrustrated counterpart. Moreover, we observed that the entanglement entropy scaling is consistent with the presence of three bosonic quasiparticles in the ground state of the system. From one side, this could suggest the conjecture that the number of excitations induced by topological frustration will be equal, in a more general and complex 2D system, to the number of extensive-length TF loops in the system. However, this reasoning might be flawed due to the fact that we considered a quasi-1D system, and for larger system sizes one might instead observe a non-trivial dimensional crossover to a different, maybe unconventional, purely 2D behaviour of topologically frustrated systems. We leave the investigation of this interesting problem to a future work.


\section*{Data and code availability} \label{sec:data-availability}

The datasets are available upon the reasonable request to the authors. The simulations were partially performed using the open source Quantum TEA library~\cite{qtealeaves_v1_5_8}. 

\begin{acknowledgments}

We thank Philippe Lecheminant, Federico Becca and Siddhartha Lal for their insights on the general behavior of quantum spin ladders. The research leading to these results has received funding from the following organizations: European Union via UNIPhD programme (Horizon 2020 under Marie Skłodowska-Curie grant agreement No. 101034319 and NextGenerationEU), Italian Research Center on HPC, Big Data and Quantum Computing (NextGenerationEU Project No. CN00000013), project EuRyQa (Horizon 2020), project PASQuanS2 (Quantum Technology Flagship); Italian Ministry of University and Research (MUR) via: Quantum Frontiers (the Departments of Excellence 2023-2027); the World Class Research Infrastructure - Quantum Computing and Simulation Center (QCSC) of Padova University; Istituto Nazionale di Fisica Nucleare (INFN): iniziativa specifica IS-QUANTUM. We acknowledge computational resources from Cineca on the Leonardo machine. SBK acknowledges support from the Croatian Science Foundation (HRZZ) Projects DOK-2020-01-9938.

\end{acknowledgments}


\appendix
\section{ground state manifold at the classical point}
\label{ap:classical_gs_manifold}
In this section we will analyze the structure of the ground state manifold of the 2D Ising model on a torus. 
We denote the number of spins along each direction respectively as $L_x$ and $L_y$, corresponding to a total number of $N=L_xL_y$ of spins in the system. Then, the system is topologically frustrated along both directions. 

A ground state can be seen as an $L_y\times L_x $ matrix whose rows is made by \textit{kink states}, i.e., antiferromagnetically aligned states with one ferromagnetic defect at some lattice site. Moreover, these rows of kink states must be arranged in such a way that the number of bonds broken (i.e., the number of ferromagnetic kinks) along the vertical direction is equal to $L_x$. Therefore, if we number the rows of the matrix from $1$ to $L_y$, and call $b_i$ the number of bonds broken between row $i$ and row $i+1$ then we must have $\sum_{i=1}^{L_y} b_i=L_x$. Of course, because of the periodic boundary conditions, the $(L_y+1)$-th row is considered to be equal to the first row.
To make a visual example, consider the following situation:
\begin{center}
    \begin{blockarray}{ccccc}
        \begin{block}{ ccccc}
             \tikzmarknode{A11}{+} & - &  + & \tikzmarknode{A14}{+} & \tikzmarknode{A15}{-} \\
            \tikzmarknode{A21}{+} &  + & - &  \tikzmarknode{A24}{+} &\tikzmarknode{A25}{-} \\
        \end{block}
    \end{blockarray}
\end{center}
\begin{tikzpicture}[remember picture,overlay]
 \draw let \p1=($(A21)-(A11)$),\n1={atan2(\y1,\x1)} in 
 node[rotate fit=\n1,fit=(A11) (A21),draw,rounded corners,inner sep=2pt]{};
 \draw let \p1=($(A24)-(A14)$),\n1={atan2(\y1,\x1)} in 
 node[rotate fit=\n1,fit=(A14) (A24),draw,rounded corners,inner sep=2pt]{};
 \draw let \p1=($(A25)-(A15)$),\n1={atan2(\y1,\x1)} in 
 node[rotate fit=\n1,fit=(A15) (A25),draw,rounded corners,inner sep=2pt]{};
\end{tikzpicture}

If we denote by $\ket{i,+(-)}$ a kink state at position $i\in \{1,5 \}$ and spin up (down), then in the above example we have that the first row is represented by the state $\ket{3,+}$ while the second is represented by $\ket{1,+}$. The number of broken bonds between the two rows along the vertical direction is $b=3$, since we can count three ferromagnetic kinks between the two layers. Following this reasoning, we can describe all the possible ground states as a collection of $L_y$ numbers $(b_1, b_2, \dots, b_{L_y})$ subject to the condition $\sum_{i=1}^{L_y} b_i=L_x$, where each $b_i$ is counting how many vertical bonds are broken between row $i$ and $i+1$. 

Our goal is therefore to understand how to construct a ground state configuration given 
an initial row described by the kink state $\ket{i,+}$ and a string $(b_1, b_2, \dots, b_{L_y})$. At this scope, we will use the following construction rules: 
\begin{itemize}
    \item let $\hat{\Pi}$ be the reflection operator, such that
    \begin{equation}
    \hat{\Pi}\ket{i,+}=\ket{i,-}. 
    \end{equation}
    It is easy to understand that the number of bonds broken between the layers $\ket{i,+}$ and $\ket{i,-}$ is $b=0$, since by construction all the spins that are vertically paired have opposite magnetization because of the action of the parity operator; 

    \item let $\hat{T}_n$ be the horizontal translation operator, such that
    \begin{equation}
    \hat{T}_n\ket{i,+}=\ket{i+n,+}, 
    \end{equation}
    for $n=1,\dots,L_x$. Of course, because of the lattice periodicity, we have that $\hat{T}_{L_x}=\mathds{1}$. It can be checked that the number of broken bonds between the states $\ket{i,+}$ and $\ket{i+n,+}$ is
    \begin{equation}
        b=\begin{cases}
            n\text, & n\text{ odd}; \\
            L_x-n\text, & n\text{ even}
        \end{cases}.
    \end{equation}
    Therefore, the number of vertical broken bonds when we apply a simple translation is always odd. 
    
    \item In order to break an even number of bonds we may combine a spatial translation and a spin reflection. The number of broken bonds between the states $\ket{i,+}$ and $\ket{i+n,-}$ is
    \begin{equation}
        b=\begin{cases}
            n\text, & n\text{ even}; \\
            L_x-n\text, & n\text{ odd}
        \end{cases}.
    \end{equation}
\end{itemize}
Therefore, given an initial state $\ket{i,+}$ for the first row and a string $(b_1, b_2, \dots, b_{L_y})$ we can construct each successive row of the matrix by applying an operator that breaks a certain amount of bonds according to the rules above described. In such a way, we have that the $l$-th row of the system can be written as $\prod_{j=1}^l \hat{O}_j \ket{i,+}$. Moreover, because of the boundary conditions we have that $\prod_{j=1}^{L_y} \hat{O}_j \ket{i,+}=\ket{i,+}$, and therefore the condition
\begin{equation}
    \prod_{j=1}^{L_y} \hat{O}_j=\mathds{1}
\end{equation}
must be satisfied.

To better understand this constructive method, let us make a practical example for a $7\times7$ system. 
Suppose that we take $\ket{1,+}$ as initial row and we are given the string $(0,2,1,3,0,0,1)$ for the broken bonds. Then, a ground state configuration can be constructed as follows
\begin{center}
    \begin{blockarray}{cccccccc}
        \begin{block}{ ccccccc|c}
            +&+&-&+&-&+&-& $\hat{\Pi}$\\
            -&-&+&-&+&-&+& $\hat{T}_5\hat{\Pi} $\\
            -&+&-&+&-&+&+& $\hat{T}_6$\\
            +&-&+&-&+&+&-& $\hat{T}_4$\\
            -&+&+&-&+&-&+& $\hat{\Pi}$\\
            +&-&-&+&-&+&-& $\hat{\Pi}$\\
            -&+&+&-&+&-&+& $\hat{T}_6$\\
        \end{block}
    \end{blockarray}
\end{center}
In the rightmost column we have written the transformation that must be applied to that row in order to obtain the next row.
One can indeed check that all the conditions on the number of bonds on the composition of the operators are satisfied. However, we could have chosen another configuration as well. Indeed, the first time we applied an operation to break two bonds, i.e., $\hat{T}_5\hat{\Pi}$, we could have also chosen to break the two bonds using the operator $\hat{T}_2\hat{\Pi}$. This would have led us to a different configuration, i.e.,
\begin{center}
    \begin{blockarray}{cccccccc}
        \begin{block}{ ccccccc|c}
            +&+&-&+&-&+&-& $\hat{\Pi}$\\
            -&-&+&-&+&-&+& $\hat{T}_2\hat{\Pi} $\\
            +&-&+&+&-&+&-& $\hat{T}_1$\\
            -&+&-&+&+&-&+& $\hat{T}_3$\\
            +&-&+&-&+&-&+& $\hat{\Pi}$\\
            -&+&-&+&-&+&-& $\hat{\Pi}$\\
            +&-&+&-&+&-&+& $\hat{T}_1$\\
        \end{block}
    \end{blockarray}
\end{center}
In general, for any string, we therefore have two possible choices for the set of transformations to be applied. This can be understood in the following way: since the sequence of operators must satisfy the condition $\prod_{j=1}^{L_y} \hat{O}_j=\mathbb{I}$, this means that the sum of the indices of the translation operators appearing in the product must be equal to an integer multiple of $L_x$. Because of the constructive rule, we see that a number of bonds $b_i$ can be broken using either $T_{b_i}$ or $T_{L_x-b_i}$, opportunely combined with reflection operators if needed. Therefore, the two possible choices that we have, given a bond string $(b_1, b_2, \dots, b_{L_y})$, are choosing all the translation of the form $T_{b_i}$ or all of the form $T_{L_x-b_i}$. In this way, the sum of the indices will always be an integer multiple of $L_x$, as required. Therefore, for all the strings involving translation operators in the construction of the ground state we have a total multiplicity of $4L_x$. The factor $4$ comes from $\mathbb{Z}_2$ symmetry and cyclic condition on the translations, while the factor $L_x$ comes from translational invariance (given a ground state matrix, we can perform $L_x$ cyclic permutations of the columns which are still ground states, or, similarly, we have $L_x$ possible choices for the position of the kink in the first row). For the strings of the form $(0,0,\ldots,0,L_x)$ and their permutations, translation operations are not involved in the construction of the ground state, therefore the multiplicity is just $2L_x$. 

The method presented above gives a constructive algorithm to build the ground states of the topologically frustrated 2D Ising model. Moreover, it allows us to evaluate the exact ground state degeneracy at the classical point.

Indeed, the total number of strings of the form $(b_1, b_2, \dots, b_{L_y})$ with $\sum_{j=1}^{L_y}b_j=L_x$ can be seen as the number of ways in which we can choose $L_y$ integer numbers between $0$ and $L_x$ such that their sum is equal to $L_x$. This problem can be reformulated as finding the $L_x$-th order Taylor coefficient of the function \begin{equation}
    f(x)=\bigg(\sum_{j=0}^{L_x} x^j\bigg)^{L_y}.
\end{equation}
Since we are interested in the Taylor coefficient about $x=0$ we can sum the partial geometric series and obtain
\begin{equation}
\begin{split}
    f(x)&=\frac{(1-x^{L_x+1})^{L_y}}{(1-x)^{L_y}} \\
    &=(1-x^{L_x+1})^{L_y}\sum_{n=0}^{\infty}(-1)^{L_y}\binom{-L_y}{n}x^n. \\
\end{split}
\end{equation}
The $L_x$-th order coefficient is therefore given by $(-1)^{L_y}\binom{-L_y}{L_x}=\binom{L_x+L_y-1}{L_x}$. Therefore, the total number of ground states will be 
\begin{equation}
4L_x\binom{L_x+L_y-1}{L_x} -2L_xL_y,    
\label{eq:degeneracy}
\end{equation}
where the subtraction accounts for the overcounting of the multiplicity of the configurations described by strings of the type $(0,0,\dots,0,L_x)$, which is only $2L_x$. 
\section{Fitting coefficients}
\label{app:fitting}

Here, we report the obtained fitting coefficients and their errors for polynomial fits in Fig.~\ref{fig:extrapolation} and Fig.~\ref{fig:QP_picture}b. In both cases, we take the logarithm of the data and recast the polynomial as a linear relation for which we compute the coefficients via least square fit. In Fig.~\ref{fig:extrapolation}, we fit to the form $\log (|h_c - h_c^*|) = a\log(L) + b$, with the obtained values for $a$ and $b$ presented in Table~\ref{tab:coeffs1}.

\begin{table}[H]
\centering
\begin{tabular}{c|c|c}
    & FM                 & AFM            \\ \hline
$a$ & -1.740 $\pm$ 0.009 & -4.2 $\pm$ 0.2 \\
$b$ & 0.96 $\pm$ 0.01    & 4.5 $\pm$ 0.3 
\end{tabular}
\caption{Fitting coefficients for Fig.~\ref{fig:extrapolation}.}
\label{tab:coeffs1}
\end{table}

\noindent In Fig.~\ref{fig:QP_picture}, we fit to the form $\log (1-R) = a\log(3L) + b$, with the obtained values for $a$ and $b$ presented in Table~\ref{tab:coeffs2}.

\begin{table}[H]
\centering
\begin{tabular}{c|c|c|c}
    & $x=1/2$           & $x=1/3$           & $x=1/4$           \\ \hline
$a$ & -0.902 $\pm$ 0.005 & -0.899 $\pm$ 0.004 & -0.891 $\pm$ 0.005 \\
$b$ & 1.556 $\pm$ 0.024 & 1.495 $\pm$ 0.018 & 1.454 $\pm$ 0.027
\end{tabular}
\caption{Fitting coefficients for Fig.~\ref{fig:QP_picture}b.}
\label{tab:coeffs2}
\end{table}

\section{Comparison with different quasiparticle ansätze}
\label{ap:ansatze}
\begin{figure}[H]
    \includegraphics[width=\columnwidth]{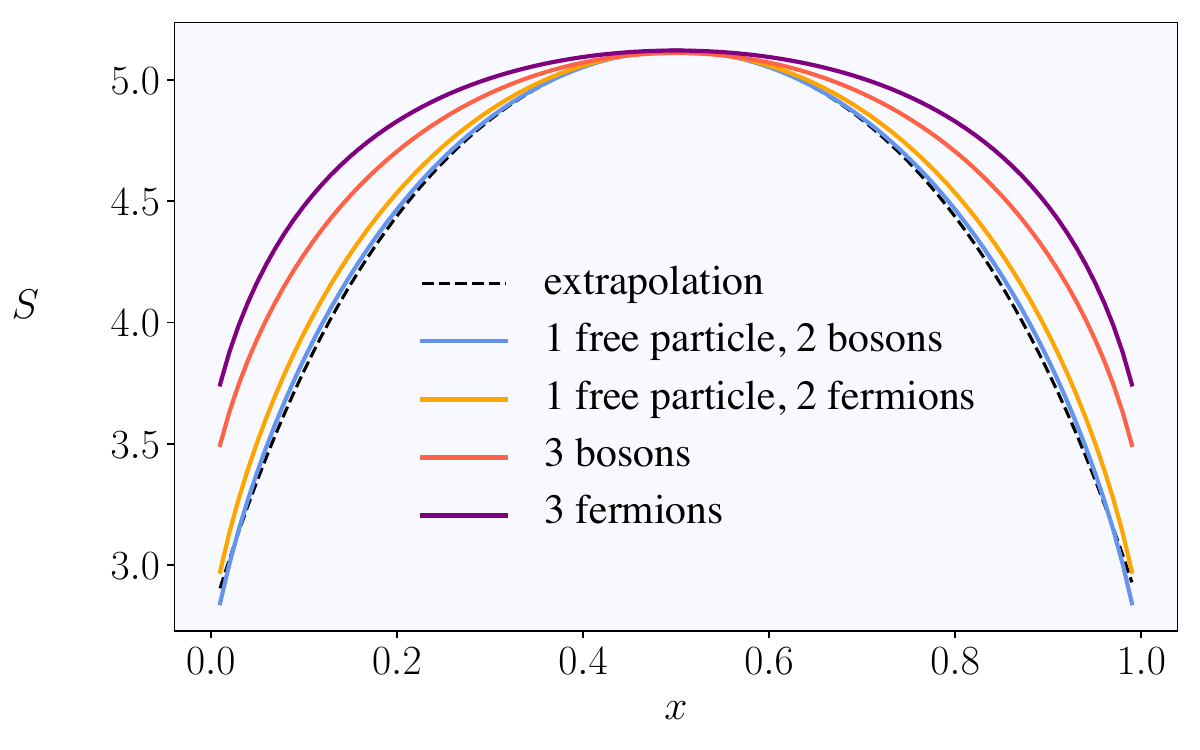} 
  \caption{\textbf{Different ansätze comparison}. The colored solid lines represent entanglement entropy $S$ as a function of the bipartition size $x$ for different quasiparticle ansatz. The dashed line represents the thermodynamic limit extrapolation of our numerical data, with the largest system size $L=71$.}
  \label{fig:all-ansatze}
\end{figure}

Here, for a matter of completeness, we provide a comparison between our numerics for the frustrated Ising ladder and different ansätze for the quasiparticle picture of entanglement. In particular, we measure the entanglement entropy for quasiparticle states (fermionic or bosonic) on a ring using numerically exact techniques to access the reduced density matrix of these systems, as outlined in~\cite{Berkovits2013}. As shown in Fig.~\ref{fig:all-ansatze}, the ansatz that best reproduces our results is the bosonic one presented in Eq.~\eqref{eq:QP_ansatz}.

\section{Details on the numerical method} \label{sec:numerical-method}

\begin{figure}[h!]
    \includegraphics[width=0.3\textwidth]{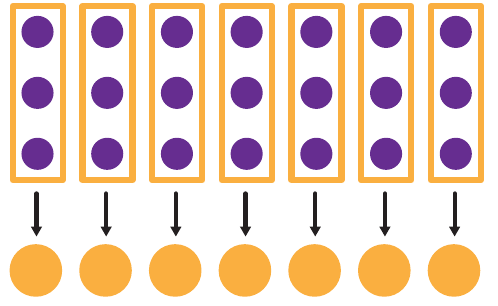} 
  \caption{\textbf{Mapping of the three-legged ladder to a one-dimensional system}. We group the sites of the same ladder rung into a single site of larger local dimension.}
  \label{fig:mapping}
\end{figure}

To find the ground states in the matrix product state (MPS) form, we run a variational DMRG algorithm~\cite{Schollwock2011, Montangero2018}. We perform the simulations of the three-legged ladder by grouping the three spins on the same rung into a single site of larger local dimension $d=8$ (Fig.~\ref{fig:mapping}). This way, the simulated model includes at most nearest-neighbour interactions.


Numerical simulation of the topologically frustrated models is extremely challenging due to the additional quasiparticle contribution to the entanglement entropy (Eq.~\ref{eq:QP_ansatz}) and large ground state (quasi)degeneracy in the bulk of the frustrated phase. In particular, unlike the one-dimensional frustrated quantum Ising model where the ground state degeneracy was linear in system size, the degeneracy in the two-dimensional case grows much more rapidly (see Eq.~\eqref{eq:degeneracy}). To escape the local minima, we find the ground states for different external fields $h$ sequentially, using the obtained state $\Psi_{\mathrm{MPS}}(h)$ as the initial guess for optimizing $\Psi_{\mathrm{MPS}}(h- \Delta h)$. The simulations are ran for a fixed bond dimension $m$, with the largest being $m=1250$ used for $L=71$.
Moreover, to find the energy gap between the ground and the first excited state in Sec.~\ref{sec:qpd}, we exploit the global $\mathds{Z}_2$ symmetry of a quantum Ising model. The first two lowest-energy states belong to different symmetry sectors. We find the lowest-energy state in each of the sectors by encoding a $\mathds{Z}_2$ symmetry into the tensor network and restricting the variational search to the manifold of states within that sector \cite{Silvi2019}. Exceptionally, for the simulations within a symmetry sector we mapped the three-legged ladder lattice to a 1D system using zig-zag mapping instead of grouping the sites as in Fig.~\ref{fig:mapping}.

\bibliography{refs}

\end{document}